\documentclass[
  aps,
  prb,
  preprint,
  superscriptaddress,
  longbibliography,
  floatfix,
]{revtex4-2}
 
\usepackage[utf8]{inputenc}
\usepackage[T1]{fontenc}
\usepackage{color}
\usepackage{graphicx}
\usepackage{dcolumn}
\usepackage{bm}
\usepackage[mathlines]{lineno}
 
\usepackage{etoolbox}
\usepackage{colortbl}
 
\usepackage{ulem}
\usepackage{amsmath}
\usepackage{amssymb}
\usepackage{xcolor} 
\usepackage{hyperref} 
\DeclareUnicodeCharacter{3B4}{$\delta$}
\DeclareUnicodeCharacter{2212}{-}

\begin{document}
 
\author{Rahul Sharma}
\altaffiliation{These authors contributed equally to this work.}
\affiliation{Université Grenoble Alpes, CEA, CNRS, IRIG-SPINTEC, 38000 Grenoble, France}
 
\author{Soumya Mukherjee}
\altaffiliation{These authors contributed equally to this work.}
\affiliation{Laboratoire de Physique de l’Ecole normale sup\'{e}rieure, ENS, Universit\'{e} PSL, CNRS, Sorbonne Universit\'{e}, Universit\'{e} de Paris, 24 rue Lhomond, 75005 Paris, France}
 
\author{Fatima Ibrahim}
\altaffiliation{These authors contributed equally to this work.}
\affiliation{Université Grenoble Alpes, CEA, CNRS, IRIG-SPINTEC, 38000 Grenoble, France}
 
\author{Gaétan Verdierre} 
\altaffiliation{These authors contributed equally to this work.}
\affiliation{Université Grenoble Alpes, CEA, CNRS, IRIG-SPINTEC, 38000 Grenoble, France}
 
\author{Libor Voj\'{a}\v{c}ek}
\altaffiliation{These authors contributed equally to this work.}
\affiliation{PSI Center for Scientific Computing, Theory and Data, Paul Scherrer Institute, CH-5232 Villigen-PSI, Switzerland}
 
\author{Martin Mi\v{c}ica}
\affiliation{Laboratoire de Physique de l’Ecole normale sup\'{e}rieure, ENS, Universit\'{e} PSL, CNRS, Sorbonne Universit\'{e}, Universit\'{e} de Paris, 24 rue Lhomond, 75005 Paris, France}
 
\author{Sylvain Massabeau}
\affiliation{Laboratoire Albert Fert, CNRS, Thales,
Universit\'e Paris-Saclay, F-91767 Palaiseau, France}
 
\author{Oliver Paull}
\affiliation{Laboratoire Albert Fert, CNRS, Thales,
Universit\'e Paris-Saclay, F-91767 Palaiseau, France}
 
\author{Vincent Polewczyk}
\affiliation{Université Paris-Saclay, UVSQ, CNRS, GEMaC, 78000, Versailles, France}
 
\author{Nicola Marzari}
\affiliation{PSI Center for Scientific Computing, Theory and Data, Paul Scherrer Institute, CH-5232 Villigen-PSI, Switzerland}
 
\author{Alain Marty}
\author{Isabelle Gomes de Moraes}
\author{Frédéric Bonell}
\affiliation{Université Grenoble Alpes, CEA, CNRS, IRIG-SPINTEC, 38000 Grenoble, France}
 
\author{Juliette Mangeney}
\author{Jerôme Tignon}
\author{Gauthier Krizman}
\affiliation{Laboratoire de Physique de l’Ecole normale sup\'{e}rieure, ENS, Universit\'{e} PSL, CNRS, Sorbonne Universit\'{e}, Universit\'{e} de Paris, 24 rue Lhomond, 75005 Paris, France}
 
\author{Anupam Jana}
\author{Jun Fujii}
\author{Ivana Vobornik}
\affiliation{CNR-IOM Laboratorio TASC, I-34149 Trieste, Italy}
 
\author{Federico Mazzola}
\affiliation{Department of Physics and Astronomy ‘Galileo Galilei’, University of Padova, Padova, Italy}
 
\author{Jing Li}
\author{Leticia Melo Costa}
\author{Olivier Renault}
\affiliation{Université Grenoble Alpes, CEA-LETI, 38000 Grenoble, France}
 
\author{Adrien Michon}
\affiliation{Université Côte d’Azur, CNRS, CRHEA, rue Bernard Grégory, Valbonne, France}
 
\author{Henri Jaffr\`es}
\author{Jean-Marie George}
\affiliation{Laboratoire Albert Fert, CNRS, Thales,
Universit\'e Paris-Saclay, F-91767 Palaiseau, France}
 
\author{Mairbek Chshiev}
\affiliation{Université Grenoble Alpes, CEA, CNRS, IRIG-SPINTEC, 38000 Grenoble, France}
 
\author{Sukhdeep Dhillon}
\email{Sukhdeep.Dhillon@ens.fr}
\affiliation{Laboratoire de Physique de l’Ecole normale sup\'{e}rieure, ENS, Universit\'{e} PSL, CNRS, Sorbonne Universit\'{e}, Universit\'{e} de Paris, 24 rue Lhomond, 75005 Paris, France}
 
\author{Matthieu Jamet}
\email{matthieu.jamet@cea.fr}
\affiliation{Université Grenoble Alpes, CEA, CNRS, IRIG-SPINTEC, 38000 Grenoble, France}

\title{Rashba engineering at van der Waals interfaces}



\begin{abstract}
Two-dimensional transition metal dichalcogenide (TMD) interfaces offer a versatile platform for studying emergent quantum phenomena and enabling novel device functionalities. When distinct TMD monolayers are stacked vertically or laterally stitched, their interfaces can exhibit unique electronic band alignments, giving rise to long-lived interlayer excitons, charge transfer effects, and moiré superlattices with correlated states. Here, we demonstrate that the interface between a large variety of two different epitaxially grown TMD monolayers controls the intensity and sign of the Rashba spin splitting, which is probed using THz spintronic emission. Optimized TMD heterobilayers, such as HfSe$_2$/PtSe$_2$, show enhanced THz emission that surpass the spin-to-charge conversion efficiency of bulk TMDs, confirming the presence of Rashba states with large spin splitting at the interface. By combining spin- and angle-resolved photoemission spectroscopy with density functional theory, we reveal that the electronic hybridization between the two different TMD monolayers gives rise to extended in-gap states with strong Rashba spin-orbit coupling. The choice of TMD layers enables to engineer the sign and strength of spin-to-charge conversion in van der Waals heterobilayers opening up perspectives to build efficient and tunable THz spintronic emitters. 
\end{abstract}

\maketitle

\section*{Introduction}

Spin-to-charge conversion (SCC) is at the core of several spintronic concepts and devices \cite{Manipatruni2019,Nguyen2024}. Efficient SCC, primarily governed by mechanisms such as the inverse spin Hall effect (ISHE) \cite{Dyakonov1971} and the inverse Rashba–Edelstein effect (IREE) \cite{Bychkov1984}, enables the direct conversion of pure spin currents into detectable charge currents resulting in THz electromagnetic radiation \cite{Ando2011, Rojas-Sanchez2013,Seifert2016,Jungfleisch2018}. In two dimensional (2D) materials like TMDs, SCC proceeds from the IREE which requires materials with strong spin-orbit coupling and broken crystalline or structural symmetry \cite{Yang2022,Galceran2021, Abdukayumov2024,Abdukayumov2025,Cheng2019,Micica2026}.
In the monolayer form, TMDs of the family MX$_2$ with M being a transition metal (Mo, W, Pt...) and X a chalcogen atom (S, Se or Te) exhibit crystalline symmetries preventing the existence of in-plane spin texture and hence SCC and THz spintronic emission. Indeed, 1H TMDs show mirror symmetry and 1T TMDs inversion symmetry \cite{Manzeli2017}. Consequently only the application of an external electric field \cite{Massabeau2025} or structural asymmetry enable the existence of Rashba coupling. Structural symmetry breaking can arise from the interaction with the substrate through proximity effects such as charge transfer or electronic hybridization \cite{Dau2018,Abdukayumov2024}. On the other hand, van der Waals (vdW) TMD heterostructures provide a versatile platform for tailoring spin–orbit fields through interlayer hybridization, phase engineering, and symmetry breaking. In TMD homobilayers, the most stable 2H and 1T stacking preserve crystalline symmetry preventing the formation of Rashba coupling whereas heterobilayers composed of distinct TMD monolayers (of 1H or 1T phases) fulfil the symmetry breaking condition. In such TMD heterobilayers, the formation of Rashba states basically depends on the degree of electronic hybridization between the two layers (existence of a 2D electron or hole gas in the vdW gap) and spin-orbit coupling strength. They exhibit spin–momentum locking and serve as efficient channels for ultrafast SCC. Despite significant progress, the direct experimental correlation between heterobilayer electronic properties (interlayer hybridization, spin-orbit interaction, phase combination or charge transfer) and SCC governing spintronic THz emission efficiency remains largely unexplored \cite{Abdukayumov2025}. 

In this work, we report the controlled epitaxial growth of a large range of high-quality single-crystalline semiconducting TMD heterobilayers with high spin-orbit coupling and various symmetries (\(\mathrm{HfSe_2/WSe_2}\), \(\mathrm{HfSe_2/PtSe_2}\), \(\mathrm{PtSe_2/WSe_2}\), \(\mathrm{WSe_2/MoSe_2}\), \(\mathrm{HfSe_2/MoSe_2}\) and \(\mathrm{PtSe_2/MoSe_2}\)) on graphene/SiC substrates using MBE and investigate SCC and their spintronic THz emission properties. We demonstrate a direct correlation between interlayer hybridization quantified via density functional theory (DFT) and spin-resolved angle-resolved photoemission spectroscopy (spin-ARPES) and experimentally measured spintronic THz-emission signals. Our combined \textit{ab initio} calculations and spin-ARPES measurements reveal the existence of an hybridized \textit{sombrero}-like valence band exhibiting Rashba-type spin splitting. This in-gap hybridized valence band corresponds to extended hole states localized at the interface between the two TMD layers, eventually forming a 2D hole gas. The resulting interface spin texture enables robust THz spintronic emission by the IREE. In HfSe$_2$-based heterobilayers, SCC by the IREE in bilayers even exceeds SCC by the ISHE in HfSe$_2$ multilayers. Furthermore, we qualitatively establish that the SCC efficiency depends on the crystal phase combination (1T/1T, 1T/1H or 1H/1H) and scales with the degree of electronic hybridization, spin-orbit interaction and asymmetry (difference in work functions).  These results are supported by a quantitative analysis taking into account the detailed calculated band structure, spin splitting and hole group velocity. Our findings provide a comprehensive framework for rationally selecting TMD phases, maximizing interlayer hybridization and for engineering high-performance THz spintronic emitters, opening new avenues in ultrafast spin–orbitronics and THz technology.


\section*{Results and discussion}

\subsection*{Material growth and characterization}

Different TMD heterobilayers, including HfSe$_2$, WSe$_2$, MoSe$_2$, and PtSe$_2$, were epitaxially grown on graphene/6H-SiC(0001) substrates using an in-house MBE system (see Materials and Methods)\cite{Dau2018,Vergnaud2020}.
 
\begin{figure}
    \centering
    \includegraphics[width=\linewidth]{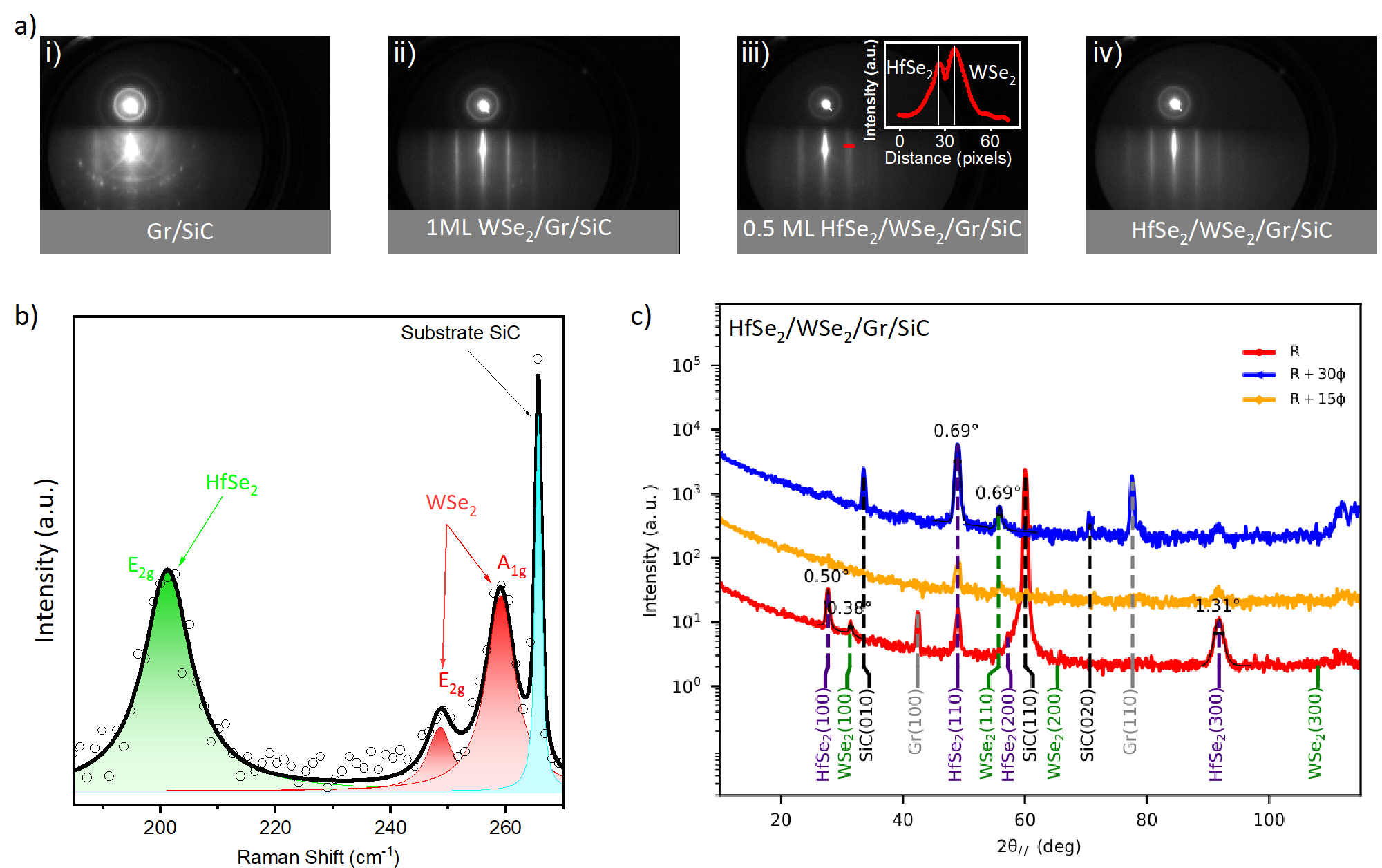}
    \caption{\textbf{Growth and Characterization of the Heterobilayers.} (a) RHEED images of (i) Gr/SiC Substrate;(ii)  WSe$_2$ monolayer; (iii) 0.5 ML HfSe$_2$ on WSe$_2$ monolayer (Inset: line profile showing the presence of two different RHEED patterns) and (iv) HfSe$_2$/WSe$_2$ heterostructure along [100]. (b) Raman Spectra from HfSe$_2$/WSe$_2$ heterostructure using a 633 nm laser (c) In-plane radial x-ray diffraction spectra along the (100) reciprocal space direction (R), (110) direction (R+30$\phi$) and at 15° from the (100) direction (R+15$\phi$). The deduced lattice parameters are indicated on the figure.}
    \label{fig:growth}
\end{figure}
Reflection high-energy electron diffraction (RHEED) was used to monitor the \textit{in~situ} layer-by-layer growth process. Figure~\ref{fig:growth}(a)i shows the characteristic diffraction patterns of the Gr/SiC substrate, which served as the template for all TMD bilayer growths. Figure~\ref{fig:growth}(a)ii displays the RHEED pattern of a monolayer WSe$_2$, while Figure~\ref{fig:growth}(a)iii presents the intermediate RHEED pattern obtained after depositing 0.5~ML of HfSe$_2$ on top of WSe$_2$. The inset in Figure~\ref{fig:growth}(a)iii shows the line profile (red line on the RHEED pattern), confirming the coexistence of two distinct TMD layers forming the heterostructure. Finally, Figure~\ref{fig:growth}(a)iv shows the RHEED pattern of the completed HfSe$_2$/WSe$_2$ heterostructure, which is dominated by the diffraction rods of the top HfSe$_2$ layer. The presence of sharp, continuous diffraction lines in all patterns indicates the high crystalline quality and 2D character of the resulting vdW heterostructures. After growth, the surface was capped with a protective 20 nm-thick amorphous selenium layer to prevent degradation before spin-ARPES and subsequent characterization. 
The as-grown heterobilayers were characterized by Raman spectroscopy and X-ray diffraction (XRD). Figure~\ref{fig:growth}(b) shows the Raman spectrum of the HfSe$_2$/WSe$_2$ heterostructure, clearly revealing the characteristic vibrational modes of the individual TMD layers: the in-plane E$_{2g}$ and out-of-plane A$_{1g}$ modes. Specifically, HfSe$_2$ exhibits the E$_{2g}$ mode near $\sim$200~cm$^{-1}$, while WSe$_2$ shows the E$_{2g}$ and A$_{1g}$ modes at $\sim$248~cm$^{-1}$ and $\sim$259~cm$^{-1}$, respectively. A background peak from the SiC substrate is also observed at $\sim$268~cm$^{-1}$. To confirm the epitaxial relationship and single-crystalline nature of the heterostructure, in-plane XRD measurements were performed. As shown in Figure~\ref{fig:growth}(c), radial scans verify that the [100] and [110] crystal directions of HfSe$_2$ and WSe$_2$ are aligned with those of the underlying graphene demonstrating the single crystalline character of TMD layers. We find the same crystalline quality for the other heterobilayers.

\subsection*{THz measurements}

Emission THz time-domain spectroscopy (TDS) measurements were performed on thickness-dependent HfSe$_2$ layers and the various heterobilayers, including HfSe$_2$/WSe$_2$ and HfSe$_2$/PtSe$_2$, and compared with homobilayer HfSe$_2$ epitaxially grown on graphene/SiC. Experimental details can be found in Materials and Methods. Characterization of THz emission using a standard THz-TDS setup is an excellent non-contact and non-destructive method to probe SCC \cite{Seifert2016}. For all samples, a 3~nm Co layer was deposited as the ferromagnetic (FM) source, followed by a 3~nm Al capping layer that naturally oxidizes into AlO$_x$. The sharpness and high quality of the FM/TMD interface were confirmed by scanning transmission electron microscopy (STEM) while the structural and chemical integrity of the TMD layer after FM deposition were demonstrated by Raman spectroscopy and X-ray photoelectron spectroscopy (XPS) (not shown here) \cite{Abdukayumov2024}. 

\begin{figure}
    \centering
    \includegraphics[width=\linewidth]{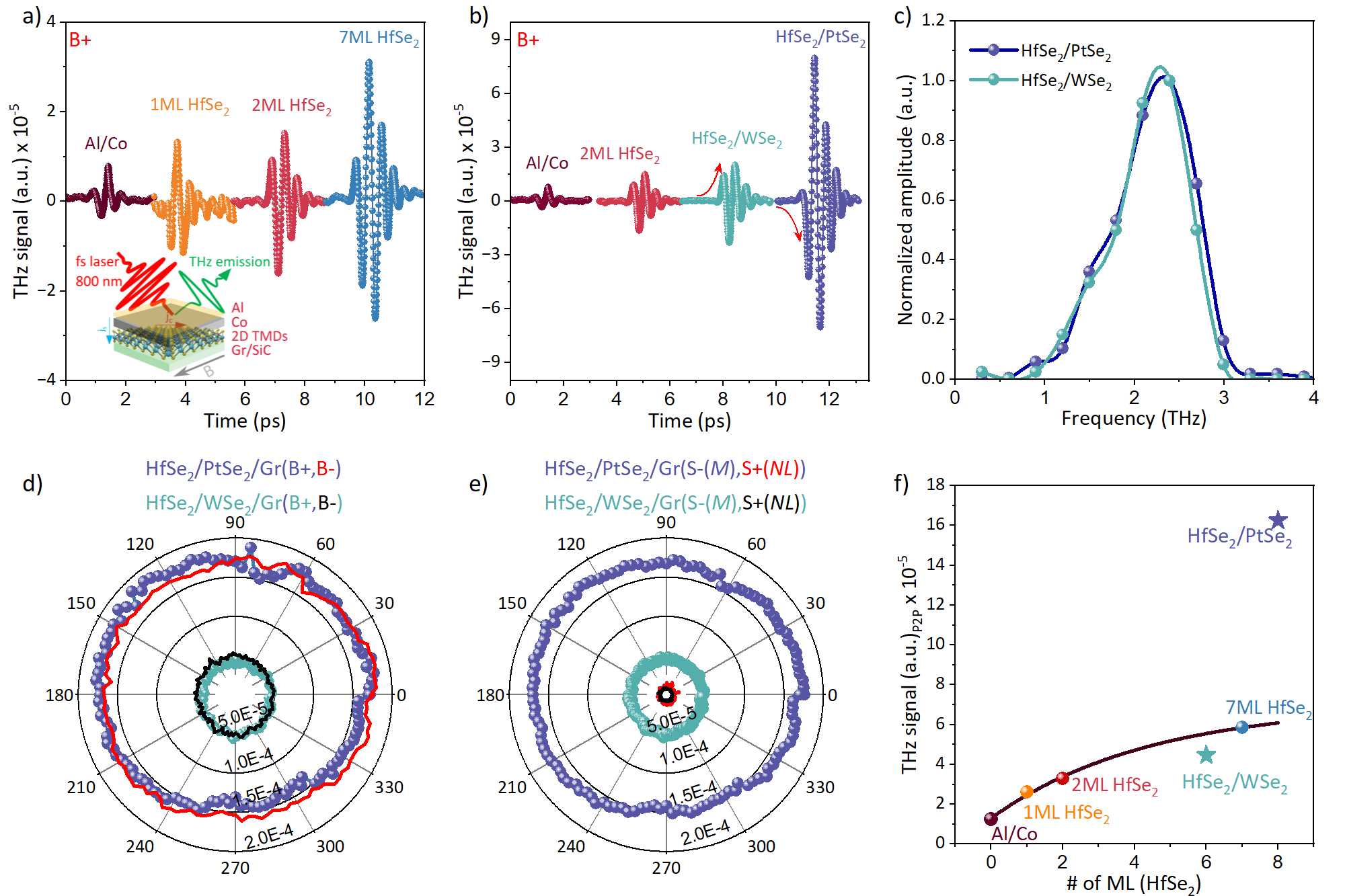}
    \caption{\textbf{THz Emission Results.} (a) Thickness dependent spintronic THz pulse emission from HfSe$_2$ (Inset: schematic of femtosecond optical pumping and THz pulse emission for TMD bilayers) (b) Comparison of THz pulse emission from homobilayers (2ML HfSe$_2$) and heterobilayers (HfSe$_2$/WSe$_2$, PtSe$_2$/WSe$_2$, and HfSe$_2$/PtSe$_2$). The red arrows indicate the opposite phase of HfSe$_2$/WSe$_2$ and HfSe$_2$/PtSe$_2$ systems. In (a) and (b), the THz waveforms are shifted horizontally along the time axis for clarity. (c) Fourier transform of THz emission from  HfSe$_2$/WSe$_2$, and HfSe$_2$/PtSe$_2$ respectively. (d) Angular dependence of the THz peak amplitude for opposite magnetic fields, (e) magnetic and non-magnetic contributions for HfSe$_2$/WSe$_2$, and HfSe$_2$/PtSe$_2$ at fixed magnetic field. (f) THz signal amplitude as a function of HfSe$_2$ thickness (orange dots). The signals of HfSe$_2$-based heterobilayers are also reported for comparison (stars).}
    \label{fig:THz}
\end{figure}

As schematically illustrated in the inset of Figure~\ref{fig:THz}(a), THz emission was measured in reflection geometry using a 800~nm (1.55~eV) pump laser that generates a spin current in the Co layers that is converted into an ultrafast current in the 2D layers. The samples were excited from the FM side in all cases. The THz electric field amplitude from HfSe$_2$ shows a clear enhancement with increasing thickness compared to monolayer HfSe$_2$. The data were fitted using the inverse spin Hall effect (ISHE) model to extract the spin diffusion length ($l_{sf}$) in pristine HfSe$_2$. The fitting results confirm a dominant ISHE contribution, indicating minimal inverse Rashba–Edelstein effect (IREE) at the HfSe$_2$/graphene interface and no evolution of the band structure of HfSe$_2$ with the number of layers as shown experimentally by momentum-resolved photoemission electron microscopy (kPEEM, see Fig. S1 of the Supp. Info.). We find $l_{sf}\approx$2.6 nm for pristine HfSe$_2$ which is within the range for typical TMDs \cite{Abdukayumov2024}. 

As shown in Figure~\ref{fig:THz}(b) (the full set of data can be found in Fig. S2 of the Supp. Info.), the heterobilayer samples produce 1.4 to 5.5 times stronger THz electric fields compared to the HfSe$_2$ homobilayer (2~ML HfSe$_2$) and nearly 3 times stronger THz electric field than 7 ML of HfSe$_2$. For detailed comparison, two representative heterobilayers are considered in the subsequent discussion: HfSe$_2$/WSe$_2$ and HfSe$_2$/PtSe$_2$. Both structures share the same top HfSe$_2$ layer but differ in the bottom TMD layer — 1T-phase PtSe$_2$ versus 1H-phase WSe$_2$ — which results in opposite THz emission polarity and a nearly fourfold difference in THz electric field amplitude for the former. Figure~\ref{fig:THz}(c) shows the normalized Fourier transforms of the THz signal from these two heterobilayers, revealing broadband emission in the 0–3~THz range (limited by the detection crystal-see Methods), comparable to state-of-the-art THz spintronic emitters.

The sample angle ($\phi$) dependence shown in Figure~\ref{fig:THz}(d) exhibits an isotropic response for both magnetic field polarities (+$B$ and -$B$) with respect to the azimuthal crystalline orientation, as expected for spin–charge conversion (SCC) mechanisms such as ISHE or IREE \cite{Rongione2023}. To separate the magnetic ($S_M$) and non-linear ($S_{NL}$) contributions from the measured THz signal ($S_{THz}$), we use \cite{Micica2026}:
\[
S_M = \frac{S_{THz}(+B) - S_{THz}(-B)}{2}, \quad
S_{NL} = \frac{S_{THz}(+B) + S_{THz}(-B)}{2}.
\]
The angular dependence in Figure~\ref{fig:THz}(e) shows a very weak non-magnetic component and an isotropic magnetic contribution, consistent with THz spintronic emission in both heterobilayer cases. Interestingly, we do not observe the usual six-fold symmetry associated with nonlinear THz emission from monolayer WSe$_2$ due to inversion symmetry breaking \cite{Micica2026}. This absence of symmetry likely arises from electronic hybridization between WSe$_2$ and HfSe$_2$, which modifies the electronic band structure and symmetry properties of the heterostructure.

\subsection*{DFT calculations and spin-ARPES Results}
\begin{figure}
    \centering
    \includegraphics[width=\linewidth]{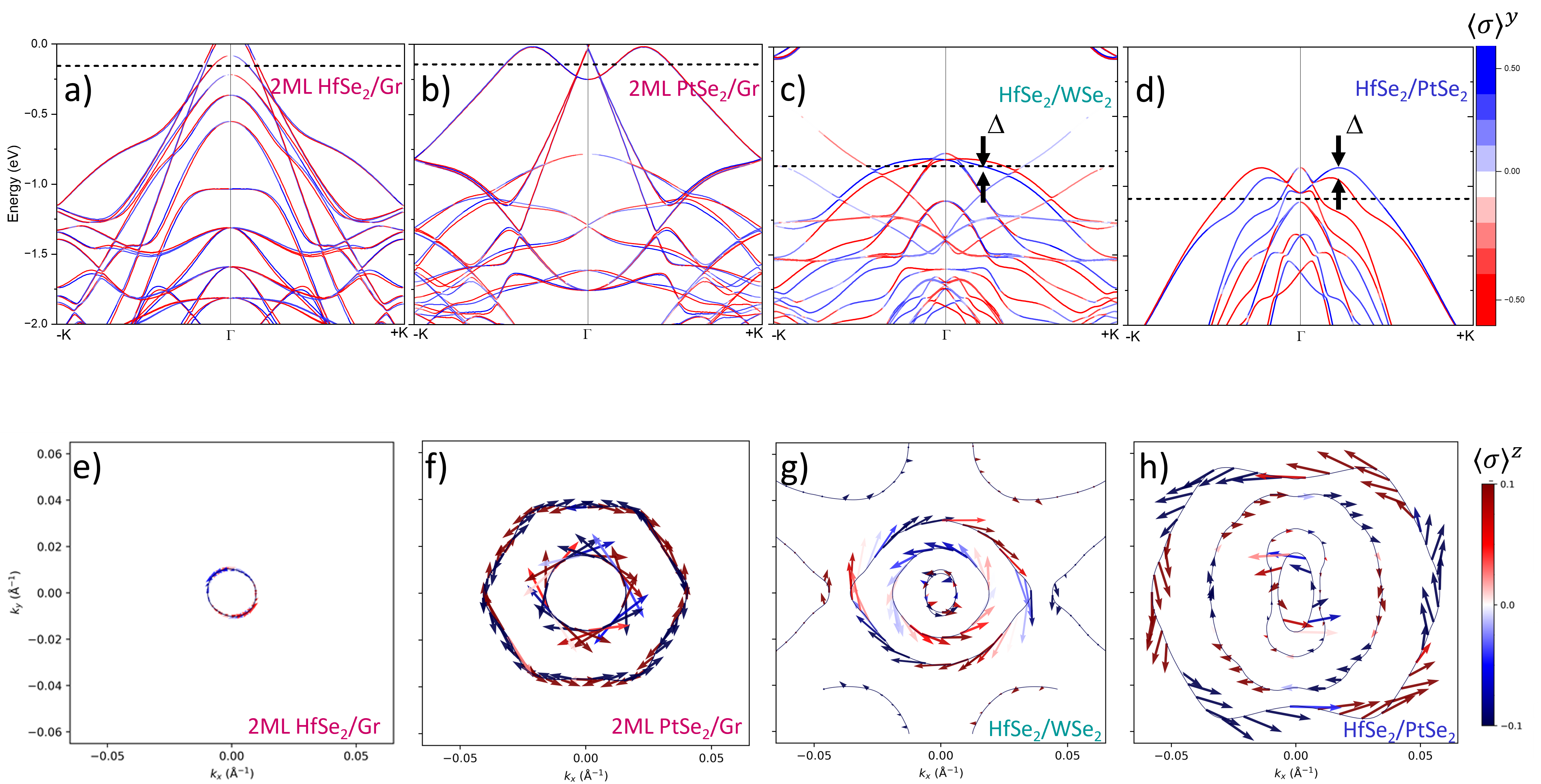}
    \caption{\textbf{DFT band-structure results.} \(S_y\)- projected band structure in the 0–2~eV range for homobilayers on graphene substrate: (a) HfSe$_2$ and (b) PtSe$_2$; and for heterobilayers: (c) HfSe$_2$/WSe$_2$ and (d) HfSe$_2$/PtSe$_2$. Panels (e–h) show spin textures in the \(k_x\)–\(k_y\) plane obtained for the top valence band at constant energy indicated by the black dashed lines in (a–d), highlighting Rashba-like spin-splitting: (e) HfSe$_2$, (f) PtSe$_2$, (g) HfSe$_2$/WSe$_2$, and (h) HfSe$_2$/PtSe$_2$. In (e) and (f), the graphene bands are not plotted.}
    \label{fig:DFT}
\end{figure}

The underlying mechanism of SCC is further understood through detailed DFT calculations. To gain microscopic insight, spin-polarized electronic band structure calculations were performed for four representative systems: two homobilayer (\(2\,\mathrm{ML}\,\mathrm{HfSe_2}\) and \(2\,\mathrm{ML}\,\mathrm{PtSe_2}\)) and two heterobilayer systems (\(\mathrm{HfSe_2}/\mathrm{WSe_2}\) and \(\mathrm{HfSe_2}/\mathrm{PtSe_2}\)). The computed band dispersions for these systems are shown in Figure~\ref{fig:DFT}(a-d). Here and in the following, we only consider the valence bands since the conduction bands exhibit almost no spin polarization in these systems. For the homobilayer systems, the band structures do not exhibit significant Rashba-type spin-split features in the valence bands (Figure~\ref{fig:DFT}a-b) as expected from crystalline inversion symmetry in 1T stacking. This is confirmed by the constant energy contours plotted in Fig.~\ref{fig:DFT}e-f where the clockwise and counterclockwise spin contours superimpose leading to zero spin polarization. Comparing the bands of the homobilayers/Gr with their pristine counterparts (Fig. S3 of the Supp. Info.) indicates negligible hybridization. Thus, the minor spin polarization observed in certain bands can primarily be due to charge transfer and resulting interfacial dipole effects at the TMD/Gr interface, rather than intrinsic Rashba effect within the bilayer itself. In contrast, the heterobilayer configurations (Figure~\ref{fig:DFT}c-d) reveal pronounced hybridization effects. Distinct \textit{sombrero}-like dispersion bands are observed near the valence band maxima, indicative of significant interlayer coupling.

Due to symmetry breaking and spin-orbit interaction, these hybridized states manifest Rashba-like spin splitting\cite{Bychkov1984} $\Delta$ (defined as the energy difference between the maxima of the two Rashba split top valence bands in Fig.~\ref{fig:DFT}(c-d)) with magnitudes of approximately \(\sim 20\,\mathrm{meV}\) for the HfSe$_2$/WSe$_2$ heterostructure and \(\sim 87\,\mathrm{meV}\) for the HfSe$_2$/PtSe$_2$ system. Figure~\ref{fig:DFT}e-h show the spin textures in the \(k_x\)–\(k_y\) plane obtained for the top valence band at constant energy. The two outermost contours with opposite spin chiralities correspond to the \textit{sombrero}-like hybridized band. The spin textures in heterobilayers are the consequence of Rasbha coupling and strongest effect can be seen in HfSe$_2$/PtSe$_2$. Notably, the spin polarization sign is opposite in the two heterobilayers, which is attributed to the opposite interfacial electric dipole fields pointing towards (away from) HfSe$_2$ in  HfSe$_2$/WSe$_2$ (HfSe$_2$/PtSe$_2$) respectively (Fig. S4 of the Supp. Info.) and differing strengths of spin-orbit coupling (SOC) arising from the constituent TMD layers. This is in agreement with the THz emission results.



\begin{figure}
    \centering
    \includegraphics[width=\linewidth, height=0.72\textheight, keepaspectratio]{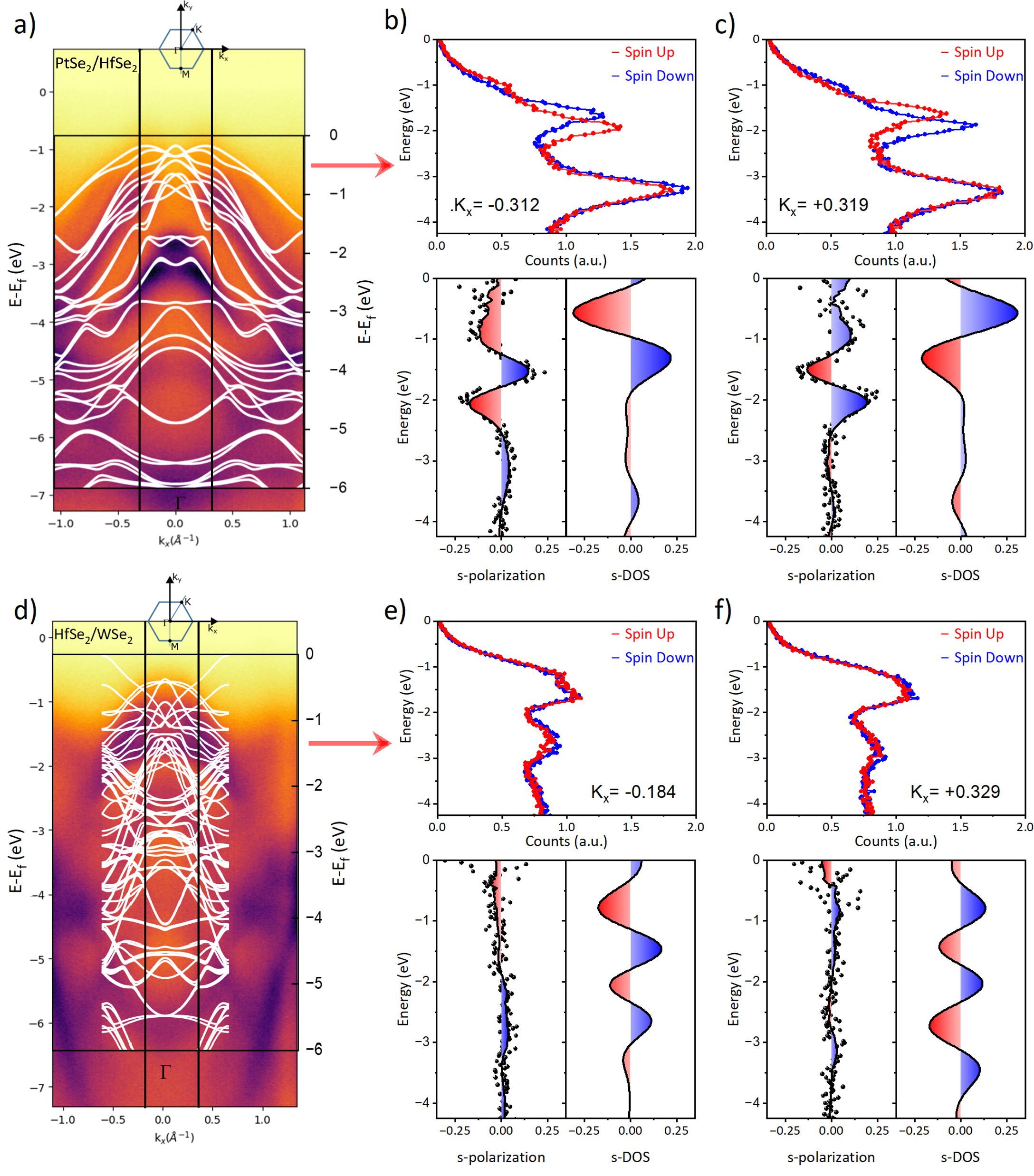}
\caption{\textbf{Spin-ARPES of the Heterobilayers.} 
(a,d) ARPES of the PtSe$_2$/HfSe$_2$/Gr  and HfSe$_2$/WSe$_2$/Gr heterostructure respectively. 
Energy sweep along the $\Gamma - K$ direction (left axis) with overlaid band structure calculated by DFT (white, right axis). The difference in energy scales between experiments and calculations is discussed in the text.
The inset shows a sketch of the first Brillouin zone. 
Spin-resolved energy distribution curves (s-EDCs) acquired at (b) $-0.312$\,\AA$^{-1}$ and (c) $+0.319$\,\AA$^{-1}$ for PtSe$_2$/HfSe$_2$, (e) $-0.184$\,\AA$^{-1}$ and (f) $+0.329$\,\AA$^{-1}$ for HfSe$_2$/WSe$_2$, corresponding to the vertical black lines in (a) and (d). The spin-polarization (s-polarization) deduced from the s-EDCs is in good agreement with the spin-density of states (s-DOS) computed by DFT.}
    \label{fig:ARPES}
\end{figure}

To further substantiate these findings, spin-ARPES measurements were performed, as shown in Figure~\ref{fig:ARPES}. The experiments were conducted at the low-energy end-station of the Advanced Photoemission Experiment (APE) beamline at the Elettra Synchrotron Radiation Facility (Trieste, Italy). The photon energy was tuned to selectively probe the valence band states, with an overall energy resolution better than \(12~\mathrm{meV}\) and a momentum resolution of approximately 0.02 \AA$^{-1}$. These high-resolution measurements enabled a precise mapping of the spin-polarized band structure near the Fermi level. The spin-integrated energy–momentum (\(E\)–\(k\)) dispersion maps recorded along the \(\Gamma - K\) high-symmetry direction of the Brillouin zone for the two heterostructures — PtSe$_2$/HfSe$_2$/Gr [Figure~\ref{fig:ARPES}(a)] and HfSe$_2$/WSe$_2$/Gr [Figure~\ref{fig:ARPES}(d)] — exhibit features that are in good agreement with the corresponding DFT-calculated band structures (as white lines), particularly the emergence of a \textit{sombrero}-like dispersion curve near the valence band maxima. This feature is absent in HfSe$_2$ homobilayer on graphene (Fig.~\ref{fig:DFT}a) and in pristine HfSe$_2$ homobilayer (see Fig. S3(b) of the Supp. Info.) and is attributed to interlayer hybridization. As demonstrated earlier from band structure calculations, these hybridized valence band states exhibit a pronounced Rashba coupling inducing spin–momentum locking.

This is further supported by the spin-resolved energy distribution curves resolved in S$_y$ [Figure~\ref{fig:ARPES}(b), (c), (e), and (f)], which clearly show opposite spin orientations at \(\pm k\), a hallmark of Rashba-type spin–momentum locking. In particular, the HfSe$_2$/PtSe$_2$/Gr heterostructure exhibits strong spin polarization signals up to $\approx$23 \% whereas the HfSe$_2$/WSe$_2$/Gr system shows comparatively a weaker spin polarization of $\approx$5 \%. The amplitude of the s-EDC are in good correspondance with the spin-polarized density of states calculated by DFT at the same momenta values k (Figure~\ref{fig:ARPES}(b), (c), (e), and (f)). Moreover, we find the same spin polarization sign for the two heterostructures. This is still in good agreement with DFT calculations since we measured the PtSe$_2$/HfSe$_2$ stack by spin-ARPES and performed DFT calculations on the reversed one (HfSe$_2$/PtSe$_2$) in Fig.~\ref{fig:DFT}. These comprehensive experimental and theoretical results together establish a robust link between the spin-resolved electronic structure, the symmetry of hybridized bands, and the resulting SCC phenomena as discussed in the following.

\subsection*{Discussion}

\begin{figure}
    \centering
    \includegraphics[width=\linewidth]{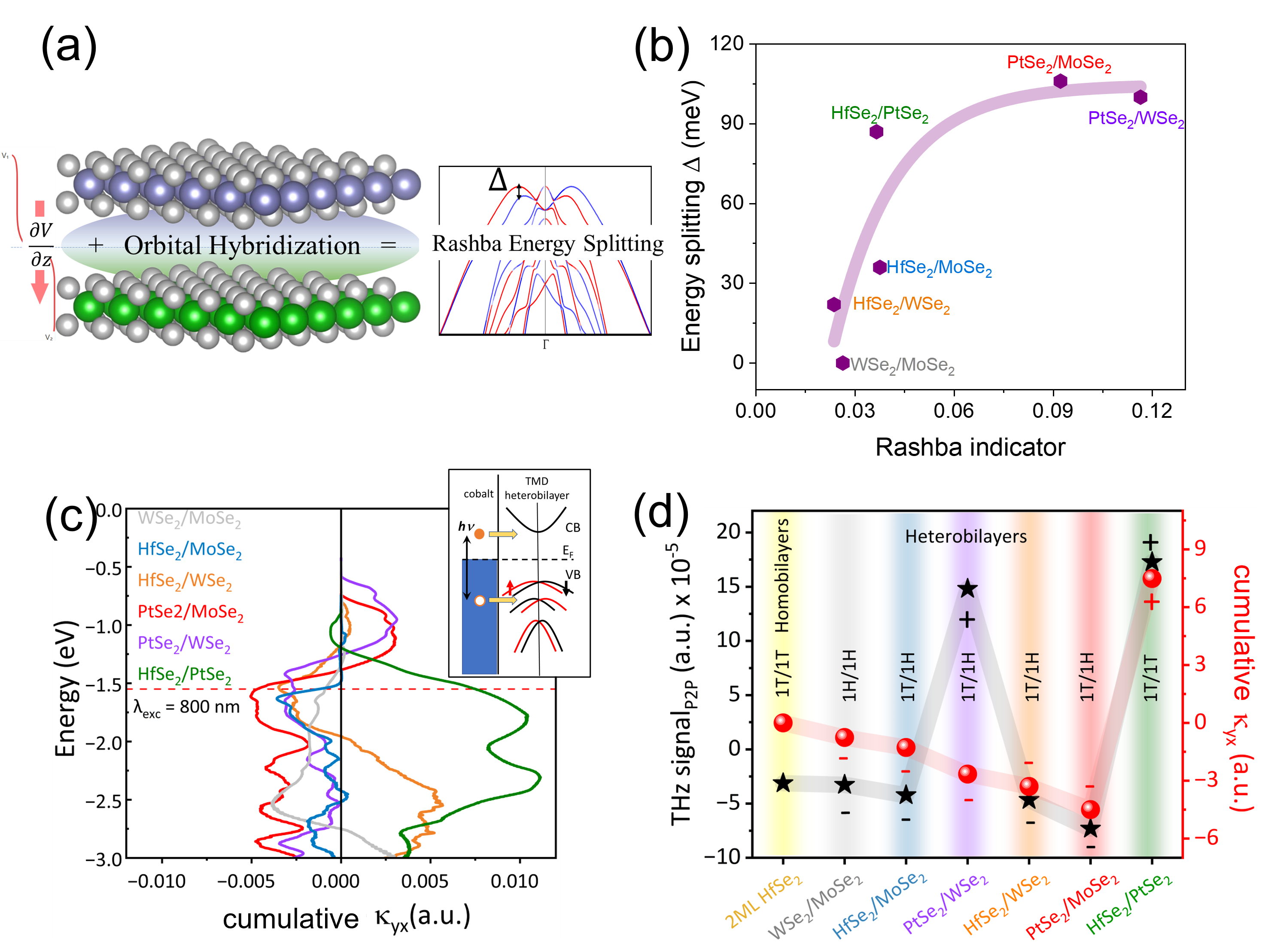}
\caption{\textbf{Trends in heterobilayer systems.} (a) Illustration of the Rashba effect in a heterobilayer with two key parameters defining the Rashba indicator: the broken symmetry represented by potential gradient ($\frac{dV}{dz}$) and the orbital hybrization across the interface. The Rashba energy splitting ($\Delta$) is calculated for the top valence band and its relation with the Rashba indicator is plotted in (b) for the different heterobilayer systems. (c) Calculated cumulative tensor $\kappa_{yx}$ for the indicated heterobilayers. The dashed horizontal line marks the excitation energy ($\lambda_{\mathrm{exc}} = 800$ nm). (d) Comparison between experiment and theory: peak-to-peak amplitude of the THz emission waveform measured under positive magnetic field $B^{+}$ (black stars) and corresponding theoretically obtained cumulative $\kappa_{yx}$ (red circles). Semi-transparent curves serve as guides to the eye. The sign of cumulative $\kappa_{yx}$ and THz response is indicated by red $(+/-)$ and black $(+/-)$ symbols placed above or below each data point, respectively. The structural phases (1T and/or 1H) of each heterobilayer are labeled above the corresponding points.}
    \label{fig:Discussion}
\end{figure}

To further elucidate the universality of the correlation between the Rashba spin splitting in TMD heterobilayers and the spin-to-charge conversion efficiency probed experimentally by THz-TDS measurements, a systematic study was performed across multiple heterobilayer systems: HfSe$_2$/WSe$_2$, HfSe$_2$/PtSe$_2$, PtSe$_2$/WSe$_2$, WSe$_2$/MoSe$_2$, HfSe$_2$/MoSe$_2$ and PtSe$_2$/MoSe$_2$ (all the THz-TDS measurements are displayed in Fig. S2 of the Supp. Info.). This is illustrated in Figure~\ref{fig:Discussion}. First, it is important to identify the ingredients leading to a large Rashba effect. Primarily, the Rashba effect is attributed to the inversion or mirror symmetry breaking at the interface associated to an electric dipole. In addition, it was demonstrated that another decisive parameter for the Rashba strength is the gradient of the Bloch-state wave-function determined by the admixture and interfacial orbital hybridization \cite{Petersen2000,Nagano2009,Krasovskii2014,Manchon2020}. On top of this, due to their inherently large SOC, heavier elements promotes a stronger Rashba splitting \cite{Shanavas2014}.       

In the following, we analyze the aforementioned parameters contributing to the Rashba effect i.e. interface asymmetry, orbital hybridization, and SOC strength in the TMD heterobilayers considered in the present work. We only focus on the hybridized \textit{sombrero}-like top valence band as being an important contributor to the THz spintronic emission of the TMD heterobilayers. For this band, the strength  of the interface hybridization is quantified as the ratio of the two TMD-projected weights denoted by the ratio of hybridization. The interface asymmetry is discerned by a potential gradient ($\frac{dV}{dz}$) and reflected as an internal electric field ($E_{internal}$) across the structure. We evaluate $E_{internal}$ by dividing the work function difference $\Delta\phi$ (Fig. S5 of the Supp. Info. shows an example for PtSe$_2$/WSe$_2$) by the thickness of the heterobilayer. This approximation is justified by the fact that the Schottky-Mott regime is usually verified in vdW multilayers \cite{LeQuang2017}. To compare the Rashba strength in the different heterobilayers, we introduce a Rashba indicator as the product of $E_{internal}$ and the ratio of orbital hybridization. The Rashba energy splitting ($\Delta$) is calculated for the top valence band as illustrated in Figure~\ref{fig:Discussion}(a) and its relation with the Rashba indicator is plotted for the different heterobilayer systems in Figure~\ref{fig:Discussion}(b). The observed increase in the energy splitting with the Rashba indicator highlights the aforementioned correlation between the Rashba strength to the asymmetry and hybridization across the TMD heterobilayers. In Figure S6 of the Supporting information, the variation of the energy splitting with $E_{internal}$ and the ratio of hybridization are plotted separately. A correlation between the Rashba spin splitting and the $E_{internal}$ is found in the region where the hybridization and large SOC do not coexist (MoSe$_2$/HfSe$_2$, HfSe$_2$/WSe$_2$, and MoSe$_2$/WSe$_2$). On the other hand, from Figure~\ref{fig:Discussion}(b), a correlation is established between the degree of hybridization and the Rashba spin splitting for heterobilayers that host both strong intrinsic SOC (namely Pt atoms) and strong orbital hybridization at the interface. Thus, we can distinguish between two different regimes. In a modest SOC strength, the asymmetry is the decisive mechanism for Rashba interaction while in a strong SOC regime the orbital hybridization becomes predominant leading to a larger Rashba effect. In summary, a large Rashba effect in TMD bilayers requires: (i) interface asymmetry in 1T/1H or 1T/1T TMD layers avoiding 1H/1H structures which has a limited structural asymmetry manifested in the weak in plane spins versus out-of-plane ones due to spin-valley locking, (ii) interfacial hybridization between the orbitals of the adjacent TMDs namely in the case where (iii) orbitals of heavy elements with intrinsic large SOC, such as Pt, are involved.\\ 

However, a direct correlation between the Rashba spin splitting of the top hybridized \textit{sombrero}-like valence band calculated by DFT and THz spintronic emission is not possible because valence bands deeper in energy have to be considered since the excitation photon energy is 1.55 eV as shown in the inset of Fig.~\ref{fig:Discussion}(c). Moreover, since the Rashba-Edelstein effect generates a charge current, the group velocity of holes has to be accounted for to compare with THz emission. Hence, to be more quantitative, we express the SCC efficiency via the tensor $\kappa$ (Ref. \citenum{Vaz2019}) which is the linear response of the unit-cell spin density $\textbf{s}\equiv \kappa\bm{\epsilon}$ to an external electric field $\bm{\epsilon}$. From the solution of the Boltzmann equation within the constant relaxation time approximation\cite{Johansson2024}, we calculate the $yx$ component of the $\kappa$ tensor per unit area of each heterobilayer as:

\begin{equation}
\kappa_{yx}=\frac{-e}{A}\sum_{k}\langle \sigma\rangle_{k}^{y}\tau_{0}v_{k}^{x}\delta (E(\textbf{k})-E_{F})
\end{equation}

where $e>0$ is the elementary charge, A is the area of the unit cell, $\langle \sigma\rangle_{k}^{y}$ is the unitless spin-$y$ expectation value, $\tau_{0}=10^{-14}$ s is a typical relaxation time\cite{Kempt2024}, $v_{k}^{x}=\frac{\partial E}{\hslash \partial k_{x}}$ is the group velocity and $\delta (E(\textbf{k})-E_{F})$ is the density of states. The band energies $E(\textbf{k})$, velocities $v_{k}^{x}$ and spin expectation values $\langle \sigma\rangle_{k}^{y}$ (Ref. \citenum{Vojacek2024}) are interpolated and summed over a dense $k$-point mesh in the first Brillouin zone, including the appropriate weights of each $k$-point. The Wannier interpolation was performed using the WANNIERBERRI package\cite{Tsirkin2021} and the comparison with DFT calculations is shown in Fig. S7 of the Supporting Information.
We then define the cumulative sum of the efficiency tensor up to an energy $E$ corresponding to the excitation photon energy ($\lambda_{exc}=800$ nm) as\cite{Vaz2019}:

\begin{equation}
\kappa_{yx}^{cum.}\equiv \int_{0}^{E}\kappa_{yx}(E')dE'
\end{equation}

Comparing the calculated SCC efficiencies for the different TMD heterobilayers to the THz measurements is hindered by two main issues. First, there is an absence of a common reference energy for the different calculated heterobilayers. Second, the band gaps of the semiconducting TMDs are under-estimated within standard Perdew–Burke–Ernzerhof (PBE)  functional, consequently affecting the position of the valence band. The combination of those two issues also explains the difference in energy scale between the experimental data and calculated bands in the spin-ARPES measurements of Fig.~\ref{fig:ARPES}. To resolve the first issue, it is essential to define a common reference energy. Since in the experiments the ultrashort laser pulse excites spin polarized electron–hole pairs in the ferromagnet (Co) generating a spin current into the heterobilayer, we choose the reference energy to be the Co Fermi level. This was done by aligning the vacuum level of the TMD heterobilayer to that of Co, where the work function of Co was calculated to be 4.9 eV. On top of that, the miss-positioning of the valence band in the TMD heterobilayers was corrected by performing hybrid Heyd-Scuseria-Ernzerhof (HSE)  functional calculations \cite{Krukau2006}. Thus, in all the plotted band structures and $\kappa_{yx}^{cum.}$, the zero energy corresponds to the Co Fermi level and the position of the top valence band includes the HSE correction. 
The energy dependences of $\kappa_{yx}$ for the 6 heterobilayers are plotted in Fig.~\ref{fig:Discussion}(c), the photon excitation energy corresponds to the horizontal dashed red line. It should be noted that the \textit{sombrero}-like bands at the top of the valence band and located in the -1.25$\rightarrow$-0.5 eV energy window are giving different $\kappa_{yx}$ values than the one at the photon energy of -1.55 eV both in sign and amplitude. This is particularly true for the HfSe$_2$/PtSe$_2$ heterobilayer which is giving the maximum THz signal and thus should exhibit maximum $\kappa_{yx}$ value. Hence, deeper valence bands have indeed to be considered for a direct comparison with experimental results. The full comparison between the calculated cumulative $\kappa_{yx}$ at -1.55 eV and the THz spintronic signal is displayed in Fig.~\ref{fig:Discussion}(d). Except for the PtSe$_2$/WSe$_2$ heterobilayer, the agreement is good both in amplitude and sign proving that our model captures well the mechanisms of SCC and THz spintronic emission. The PtSe$_2$/WSe$_2$ system requires additional investigation since the difference between calculations and measurements might come from experimental issues like unexpected stacking, thicker films or incomplete coverage. The small discrepancy between calculated and experimental values for the other systems can have several origins: the hole relaxation time (prefactor of $\kappa_{yx}$) has been taken identical for all heterobilayers which represents a strong approximation since it depends on extrinsic effects (interface roughness, impurities...) and intrinsic effects like electron-phonon scattering which might be different from bilayer to bilayer. Other parameters are not accounted for in our model like the possible presence of an electric dipole at the cobalt/TMD interface, the generation of electron-hole pairs in the heterobilayers (the excitation photon energy being larger than the direct or indirect bandgap of all heterobilayers), the presence of graphene as a substrate and hole relaxation in the ferromagnet that can also differ from bilayer to bilayer depending on the energy landscape at the interface with the TMD. Even though our model convincingly depicts most of the physics at stake, more sophisticated models would be required to match experimental findings with more accuracy.

In conclusion, we demonstrate the epitaxial growth of a large variety of single-crystalline vdW heterobilayers over large areas using MBE and systematically investigate their SCC characteristics via THz spintronic emission spectroscopy. Experimental results are well explained by DFT calculations and spin transport models. This work provides a clear design strategy for engineering high-performance SCC devices: (i) selecting appropriate TMD phase combinations (\(1H/1H\), \(1T/1H\) or \(1T/1T\)), (ii) maximizing interlayer hybridization through suitable materials choice (e.g., \(\mathrm{PtSe_2}\)), and (iii) maximizing spin-orbit coupling via a proper choice of the transition metals. The findings highlight the pivotal roles of crystalline phase, interlayer coupling, and orbital hybridization in achieving efficient spin–charge interconversion and optimizing THz emission in vdW heterostructures.

\section*{Materials and Methods}

\noindent \textbf{MBE growth}

The vdW bilayers were epitaxially grown on graphene/SiC(0001) substrate which was prepared by CVD in a hydrogen atmosphere \cite{Michon2010} under conditions leading to the formation of a single layer of graphene on a buffer layer \cite{Mastropasqua2025}. First, the substrate was degassed at 800 °C for 30 minutes, then cooled down to 400 °C, the growth temperature of WSe$_2$. Tungsten deposition rate was $\approx$0.1 \AA/min and Se flux of 1$\times$10$^{-6}$ mbar. To ensure more coverage of the substrate surface, a nominal thickness of 1.2 ML WSe$_2$ was deposited, followed by an annealing step at 820 °C for 20 minutes. 1.2 ML of HfSe$_2$ was also deposited at 340 °C on top of WSe$_2$ with the Hf deposition rate of 0.3 \AA/min under Se flux of 1$\times$10$^{-6}$ mbar. Then, the whole bilayer was annealed again at 720 °C for 15 minutes. Once at room temperature, the samples were either capped with 20 nm thick of amorphous Se (removed thermally at 280°C in the analysis chamber for spin-ARPES measurements) or covered with 3 nm of Co in the same MBE chamber with the deposition rate of $\approx$0.3 \AA/s and capped with 3 nm Al evaporated at $\approx$0.3 \AA/s that protects the ferromagnet against oxidation.

\noindent \textbf{X-ray diffraction}

The grazing incidence X-ray diffraction (GIXRD) was done with a SmartLab Rigaku diffractometer equipped with a copper rotating anode beam tube (K$_{\alpha}$ = 1.54 \AA) operating at 45 kV and 200 mA. Parallel in-plane collimators of 0.5° of resolution were used both on the source and detector sides.

\noindent \textbf{THz measurements}

For THz time-domain spectroscopy (THz-TDS) measurements, the polarization of the pump pulse was vertical. The sample was placed between the poles of an electromagnet with horizontal magnetic field B$\pm$=$\pm$20~mT (enough to saturate the in-plane magnetization of Co). The THz pulse detection was performed by electro-optic sampling by 500~\textmu m thick ZnTe aligned to detect vertical polarization and laser pulse from the main laser (1030~nm, 130~fs).
The angular dependence measurements (0$^{\circ}$-60$^\circ$) were performed using a Ti:Sapphire oscillator (800~nm, 1~GHz, 15~fs) in reflection configuration (sample excited from Co side with THz collection from the same side).

\noindent \textbf{DFT calculations}

Our first-principles calculations are based on the projector-augmented wave (PAW) method \cite{Blochl1994} as implemented in the VASP package \cite{Kresse1993,Kresse1996a,Kresse1996b} using the generalized gradient approximation \cite{Perdew1996} and including spin-orbit coupling. TMD heterostructures were modeled by constructing the following supercells from the unit cells so that to minimize the lattice mismatch: 2$\times$2-PtSe$_2$/3$\times$3 graphene, 2$\times$2-HfSe$_2$/3$\times$3-graphene, $\sqrt{3}\times\sqrt{3}$ HfSe$_2$/2$\times$2 WSe$_2$, 1$\times$1 HfSe$_2$/ 1$\times$1 PtSe$_2$, $\sqrt{3}\times\sqrt{3}$ HfSe$_2$/ 2$\times$2 MoSe$_2$, $\sqrt{3}\times\sqrt{3}$ PtSe$_2$/2$\times$2 WSe$_2$, $\sqrt{3}\times\sqrt{3}$ PtSe$_2$/2$\times$2 MoSe$_2$. A sufficient vacuum layer of 20 \AA\ thickness was added to the heterostructures. The atomic coordinates were relaxed until the forces became smaller than 1 meV/\AA. A kinetic energy cutoff of 550 eV has been used for the plane-wave basis set and a $\Gamma$-centered 15$\times$15$\times$1 $k$-mesh was used to sample the first Brillouin zone. To describe correctly the interaction across the interface, van der Waals forces were used with Grimme type dispersion-corrected density functional theory-D2 \cite{Bucko2010}. The Wannier tight-binding models were constructed based on maximally localized Wannier functions \cite{Marzari1997}  using the WANNIER90 package \cite{Mostofi2014}. The WANNIER90 input files with all the relevant input parameters (the $k$-point sampling of the non-self-consistent DFT step, the initial projectors, the number of disentanglement and wannierization steps, etc.) are available on demand. The Wannier models reproduce well the band structure of the original DFT calculations as shown in Fig. S7 of the Supporting Information. 

\section*{Acknowledgements}

The authors acknowledge the support from the European Union’s Horizon 2020 research and innovation Programme under grant agreements No 881603 (Graphene Flagship), No 964735 (FET-OPEN EXTREME-IR) and No 101099552 (EIC Pathfinder PLASNANO). This research was supported by the NCCR MARVEL, a National Centre of Competence in Research, funded by the Swiss National Science Foundation (grant number 205602). Computer time was provided by the Swiss National Supercomputing Centre (CSCS) under project No. lp18 and mr33. O.P. and M.B. acknowledge financial support from ERC AdG "FRESCO" (\#833973). Spintec was supported by the FLAG-ERA grant MNEMOSYN. The French National Research Agency (ANR) is acknowledged for its support through the ANR-18-CE24-0007 MAGICVALLEY, ANR-22-CE09-0012-03 COME\_ON, ANR-20-CE24-0015 ELMAX, ANR-22-CE24-0022-01 VanaSiC and ESR/EQUIPEX+ ANR-21-ESRE-0025 2D-MAG projects. The LANEF framework (No. ANR-10-LABX-0051) is acknowledged for its support with mutualized infrastructure. Part of the calculations were performed on computational resources provided by GENCI–IDRIS (Grant 2024-A0150912036). The authors also acknowledge the France 2030 government investment plan PEPR Electronics through the ANR-22-PEEL-0011 ADICT project and PEPR SPIN through the ANR-22-EXSP-0003 TOAST, ANR-22-EXSP-0007 SPINMAT and ANR-22-EXSP-0009 SPINTHEORY projects. The authors also acknowledge Ekta Yadav for helping in THz spintronic emission measurements at Laboratoire Albert Fert, Palaiseau, France.

\vspace{1 cm}

\noindent The authors declare no competing interests.

\pagebreak
\end{document}